\documentstyle[12pt]{article}

\renewcommand{\Large}{\large}
\catcode `@=11 \@addtoreset{equation}{section} \catcode `@=12

\begin{document}

\def\be{\begin{equation}}
\def\ee{\end{equation}}
\def\ba{\begin{array}{l}}
\def\ea{\end{array}}
\def\bea{\begin{eqnarray}}
\def\eea{\end{eqnarray}}
\def\eq#1{(\ref{#1})}
\def\ch{{\rm cosh}\alpha}
\def\th{{\rm tanh}\alpha}
\def\G{\Gamma}
\def\rhot{{\tilde\rho}}
\def\V{{\cal V}}
\def\z{{\bar z}}
\def\w{{\bar w}}
\def\del{\partial}
\def\M{{\cal M}}

\renewcommand\arraystretch{1.5}

\begin{flushright}
TIFR-TH-95/61\\
hep-th/9511218
\end{flushright}
\begin{center}
\vspace{3 ex}
{\Large\bf
BLACK HOLE GEOMETRY AROUND AN ELEMENTARY BPS STRING STATE
}\\
\vspace{10 ex}
Gautam Mandal and Spenta R. Wadia \\
Tata Institute of Fundamental Research \\
Homi Bhabha Road, Bombay 400 005, INDIA \\
\vspace{1 ex}
e-mail: mandal, wadia@theory.tifr.res.in\\
\vspace{15 ex}
\bf ABSTRACT\\
\end{center}
\vspace{2 ex}
We probe the geometry around an elementary BPS (EBPS) state in
heterotic string theory compactified on a six-torus by scattering a
massless scalar off it and comparing with the corresponding experiment
in which the EBPS state is replaced by a classical extremal black hole
background satisfying the BPS condition. We find that the low energy
limit of the scattering amplitudes precisely agree if one takes the
limit $m_{\rm bh} >> m_P$. In the classical experiment, beyond a
certain frequency of the incident wave, part of the wave is found to
be absorbed by the black hole, whereas in case of the string
scattering there is a critical frequency (inelastic threshold) of the
probe beyond which the EBPS state gets excited to a higher mass
non-BPS elementary state. The classical absorption threshold matches
exactly with the inelastic threshold in the limit of maximum
degeneracy of the EBPS state of a given mass. In that limit we can
therefore identify absorption by the black hole as excitation of the
elementary string state to the next vibrational state of the string
and consequently also identify the non-BPS string state as a
non-extremal black hole.
\vfill
\eject

\setcounter{section}{-1}

\section{Introduction}

It is well-known that elementary particles heavier than the Planck
mass have Compton wavelength smaller than their classical
`Schwarzschild radius'. The suggestion that such particles may be
identified with black holes and conversely black holes may be viewed
as elementary particles has a long history \cite{black}. Stable
excitations of such masses abound in $N=2$ string compactifications,
and if the above identification is true, it might have profound effect
on our understanding of both black hole physics and string theory.
Recently there have been a number of papers
\cite{Duff,Sen,Khuri,Dabholkar,Wilczek,Susskind} investigating from
various viewpoints the correspondence between classical extremal (BPS)
black hole solutions in toroidally compactified heterotic string and
elementary BPS states in the same theory carrying the same mass and
charge as the black hole.

In this paper we probe the geometry around an elementray BPS (EBPS)
state by the following `experiments'. We choose as our probe a
massless scalar excitation of the string theory (we work with
heterotic string compactified on a six-torus throughout). We scatter
this probe off (a) the extremal black hole solution and (b) the EBPS
state in flat space carrying the same quantum numbers (mass and
charge). The idea is that if the background created by the EBPS state
is the same as that of the classical solution, the results should
agree. Here are the main results:

(i) The classical scattering calculation involves solving the wave
equation of the massless scalar (of frequency $w$) in the black hole
background and calculating the phase shifts $\delta_l$ of the partial
waves (Sec 1). The string scattering (Sec 2) involves calculating the
tree-level four-point amplitude $\langle$ EBPS-probe-probe-EBPS
$\rangle$. The low energy limit of this amplitude agrees (Sec 3) {\em
precisely} with the leading classical result if one further takes the
limit $m>> m_P$\footnote{We are working in Einstein units rather than
string units, so the natural scale in our problem is $m_P =
1/\sqrt{\alpha'}$ rather than $m_s = g_s m_P$.} in the string
calculation, where $m$ is the mass of the EBPS state (equal to the ADM
mass of the black hole).

(ii) The string scattering calculation explicitly involves internal
polarization tensors which appear in the vertex operator for the EBPS
state, whereas these do not appear in the specification of the
classical background. In the limit mentioned above any nontrivial
dependence of the string amplitude on the `hair' (internal
polarization tensors of the EBPS state) disappears (Sec 2).

(iii) In the classical calculation there exists a critical frequency
$w_{cr}$ such that when the frequency $w$ of the massless scalar wave
exceeds $w_{cr}$ the phase shifts $\delta_l$ start becoming complex
and the incoming wave gets partly absorbed by the black hole. On the
string side there exists a critical frequency $w'_{cr}$ such that for
$w\ge w'_{cr}$ the EBPS state can get excited to a non-BPS elementary
state. These two critical frequencies exactly coincide when the ratio
$|Q_L|/|Q_R| \to 0$ ($Q_L, Q_R$ are the left- and right-moving
charges) which is also the limit of maximum allowed degeneracy of the
EBPS state of a given mass. In that limit we are therefore led to a
rather simple interpretation of absorption of matter by the black hole
as simply excitation of the string state to a higher vibrational
mode. It also implies an identification of these non-BPS states with
non-extremal black hole. We remark on the issue of the de-excitation
process and Hawking radiation in the concluding section (Sec. 4).

\section{Classical Scattering}

We consider the extremal black hole solution in heterotic string
theory compactified on a six-torus \cite{Sen}. We reproduce some of
the formulae we will need here. The low energy 4D lagrangian
describing the dynamics of the massless fields is
\bea
S = {1 \over 16 \pi G_N}\! \int \!d^4 x \sqrt{- g} [ R_g
- e^{-\Phi} F^{(a)}_{\mu \nu} (LML)_{ab} F^{(b)\mu \nu} + {1
\over 8} {\rm Tr} (\del_\mu M L \del^\mu M L) + \ldots ]
\nonumber
\eea
where $G_N$ is Newton's constant and equals $\alpha'/8$. $g_{\mu
\nu}$ is the Einstein metric and indices are contracted above
with respect to it. $M$ is a $28 \times 28$ matrix-valued field
representing the massless scalars. $L$ is a $28 \times 28$ diagonal
matrix with the first 22 entries equal to $-1$ and the remaining six
equal to 1. The terms denoted by $\ldots$ do not involve $M$. The
above lagrangian possesses a black hole solution \cite{Sen} given by
\be
ds^2 \equiv g_{\mu\nu} dx^\mu dx^\nu = -\rho/\sqrt K dt^2 + \sqrt K/\rho
d\rho^2 + \rho \sqrt K d\Omega^2
\label{1.1}
\ee
where
\be
\ba
K = \rho^2 + 2 m_0 \ch\, \rho + m_0^2 \\
\th  = |Q_L|/|Q_R| \\
m_0 = \alpha' m/(4 \ch)
\ea
\label{1.2}
\ee
We denote by $Q_R, Q_L$ the charge vectors in the right-moving and
left-moving sectors (and by $|Q_{L,R}|$ their magnitudes)
respectively. $m$ is the ADM mass of the black hole.

We consider the propagation of fluctuations of the massless scalar
($\delta M_{ab}$) in such a background. The equation of motion is
given by
\be
D^\mu \del_\mu \delta M_{ab} = 0
\label{1.3}
\ee
This can be derived by putting $M_{ab} = M^{\rm cl}_{ab} + \delta
M_{ab}$ in the above lagrangian. The terms linear in $\delta M$ vanish
by classical equation of motion for the background $M^{\rm cl}$. The
remaining terms involving $\delta M$ are quadratic and lead to
\eq{1.3}. The metric appearing in the above equation is the `Einstein'
metric.  We choose the scalars as probes because their propagation
equation is particularly simple, involving only the `Einstein' metric
and no other background.

The classical scattering of $\delta M_{ab}$ off the black hole
\eq{1.1} is computed by doing the partial wave analysis of \eq{1.3}
and finding out the phase shifts as detailed below. We should remark
that \eq{1.3} is the same as the equation $L_0 = \bar L_0 = 1$ for the
vertex operator for $\delta M$ in the curved background. Consequently
a scattering calculation using curved-space sigma-model world-sheet
action and vertex operators \eq{1.3}\ for the massless scalar string
state is in principle equivalent to the following analysis.

\vspace{2 ex}

\noindent Step 1. \underbar{Partial wave analysis}:

We look for solutions of the form ({\it cf.} \cite{Schiff} )
\be
\ba
\delta M_{ab}(\rho,t, \theta, \phi) =  M_w(\rho, \theta,
\phi) \exp(-iwt)\\
M_w(\rho, \theta, \phi) = \sum_l Y_{l0} (\theta,
\phi) \, \psi_{wl}(\rho)/\rho
\ea
\label{1.4}
\ee
Substituting the above in \eq{1.3} we find that the partial waves
$\psi_{wl}$ satisfy the following equation:
\be
- \psi'' + V(\rho) \psi = 0, \qquad  V(\rho)= {l(l+1)\over \rho^2}
- w^2 {K\over \rho^2}
\label{1.5}
\ee
This equation is exactly solvable in terms of confluent hypergemotric
functions. The result is \cite{Abram}
\be
\psi(\rho) = \rho^{l'+1} \exp[iw\rho] \bigg( A M(a, 2(l'+1),
- 2 i w \rho) + B U(a, 2(l'+1) , -2 i w \rho)  \bigg)
\label{1.6}
\ee
Here
\be
\ba
l' =  - 1/2 + \sqrt{ (l + 1/2)^2 - m_0^2 w^2 }\\
a = l' + 1 - i a' , \qquad a' \equiv m_0 \ch\, w \\
\ea
\label{1.7}
\ee
In the above $A$ and $B$ are integration constants. The requirement
of finiteness of the solution at $\rho=0$ gives us the boundary
condition (for $m_0 w < 1/2$)
\be
B = 0
\label{1.8}
\ee
since $U \propto \rho^{-2 l' - 1} $ as $\rho \to 0.$
Using the $\rho\to\infty$ asymptotics of the confluent hypergeometric
functions we now get the phase shifts $\delta_l$ of the partial
waves $\psi_{wl}$:
\be
\ba
\psi_{wl}(\rho) \sim C_{wl} \sin(w\rhot + \delta_l - \pi l/2)\\
\rhot \equiv  \rho + m_0 \ch \, \ln (2w\rho) \\
\exp(2 i\delta_l) = \G(l'+1- ia')/\G(l' + 1 + ia')
\ea
\label{1.9}
\ee
The symbol $\sim$ throughout this paper will imply `asymptotically
true for large $\rho$'.

\vspace{2 ex}
\noindent Step 2. \underbar{Summing up the partial waves:}

In order to obtain the scattering amplitude we need to choose $C_{wl}$
in \eq{1.9} such that the sum over partial waves in \eq{1.4} is of the
form (as $\rho\to\infty$)
\be
M_w(\rho, \theta, \phi) \sim  \exp[iw(z - m_0 \ch \ln w(\rho-z))]
+ {f(\theta)\over \rho} \exp[iw \tilde \rho]
\label{1.10}
\ee
This is done in Appendix A. The result for $f(\theta)$ is
the following:
\be
f(\theta)= \frac{\alpha' m}{8} {\rm cosec}^2 \theta/2
\bigg[1  +  o(m_0 w, w/m_P) \bigg]
\label{1.11}
\ee
Note that the above scattering amplitude is independent of
the `polarization' $(ab)$ of $\delta M_{ab}$. We shall see that
in the string calculation also the scattering amplitude becomes
independent of the polarization of the massless scalar in
a suitable limit.

\vspace{ 2 ex}
\noindent{\bf Absorption Threshold:}
Note that the phase shifts $\delta_l$, given by \eq{1.9}, remain real
as long as $m_0 w < 1/2$, that is, as long as
\be
w \le w_{cr},  \quad  w_{cr} = 1/(2 m_0)
\label{1.12}
\ee
For $w > w_{cr}$, the phase shifts become complex, signalling
absorption. One can explicitly calculate the flux through a small
2-sphere around $\rho=0$ and show that at $w > w_{cr}$ this
becomes non-zero.

\section{String-string scattering}

We work with the heterotic string theory compactified on a
six-dimensional torus.  We will use the notation $x^\mu(z,\z),
\mu=1,\ldots,4$ for the 4 non-compact coordinates, $x_R^i(z),
i=1,\ldots, 6$ for the 6 compact right-moving coordinates and
$x_L^j(\z), j=1, \ldots, 22$ for the 22 compact left-moving
coordinates. The right-moving fermions will be denoted by
$\psi^\mu(z)$ and $\psi_R^i(z)$.  To calculate the scattering of a
massless scalar off an elementary BPS state we need the following
ingredients:

(a) Vertex operator for the massless scalar ($M_{ab}$):
\be
\ba
\V_M(\eta_R; \eta_L; k; z,\z) =
V_M(\eta_R,k, z) \bar V_M(\eta_L, \z) \exp[ik.x(z,\z)]\\
V_M(\eta_R ,k, z) =  \eta_R.(\del_z x_R + i k_\mu\psi^\mu \psi_R) \\
\bar V_M(\eta_L, \z) = \eta_L. \del_{\bar z} x_L \\
\ea
\label{2.1}
\ee

(b) Vertex operator for the elementary BPS state:

The elementary BPS states satisfy a mass formula\footnote{In the
rest of this section  we will work with the convention
$\alpha' = 2.$}
\be
m^2 = |Q_R|^2 =  |Q_L|^2 + 2 (N_L  -1)
\label{2.2}
\ee
where $Q_R, Q_L$ are the left-moving and right-moving charges.  Given
some mass $m$, $N_L$ is not fixed, so one has to consider various
cases.

\underbar{Case $N_L =1$}:  The vertex operator in this case is given by
\be
\ba
\V_B(\zeta_R; \zeta_L;k; z,\z) =
V_B(\zeta_R, k, z) \bar V_B(\zeta_L, \z) \exp[iQ_R.x_R+ iQ_L.x_L +
ik.x(z,\z)]\\
V_B(\zeta, k,z) = \zeta_R. \psi_R (z) \, e^{-\phi(z)} \\
\bar V_B(\zeta_L, \z) = \zeta_L. \del_\z x_L \\
\ea
\label{2.3}
\ee
We have chosen $V_B$ in the `$-1$' picture in order to provide
$\phi$-ghost charge $-2$ in the following four-point function. We have
also chosen polarization vectors of the EBPS state to lie only in the
internal compact directions since we want to make correspondence with
a spin-zero black hole.

We now consider scattering a probe (particle `2') off an EBPS state
(particle `1'). The final states will be denoted `4' and `3'
respectively.  We will call the initial and final polarizations
of the probe $\eta_{L,R}, \eta'_{L,R}$ and those of the EBPS state
$\zeta_{L,R}, \zeta'_{L,R}$. The four-point scattering amplitude
is given by (computed in Appendix B)
\be
\ba
\M(1,2,3,4)
\equiv \int \prod d^2 z_3 \,
\langle c\bar c(z_1) c\bar c(z_2) c \bar c(z_4)
\\
\V_B(\zeta_R; \zeta_L; k_1; z_1)
\V_M(\eta_R;\eta_L; k_2; z_2)
\V_B^{(-)}(\zeta'_R; \zeta'_L; k_3; z_3)
\V_M(\eta'_R; \eta'_L; k_4; z_4)
\rangle  \\
= \bigg[K_R^1  -{t(t+2)\over (s-m^2)(u - m^2)}K_R^2  -
{t(t+2) \over 2(s- m^2)} K_R^3 - {t(t+2)\over 2(u - m^2)} K_R^4 \bigg]
\times \\
\bigg[K_L^1 - {t(t+2) \over (s - m^2) (u - m^2)}  K_L^2 +
{t(t+2) \over (s-m^2) (s - m^2 + 2)} K_L^3  +
{t(t+2) \over (u-m^2) (u - m^2 + 2)} K_L^4 \bigg] A_1(s,t,u)
\ea
\label{2.4}
\ee
where
\bea
A_1(s,t,u) = - \pi \G(-{t\over 2}) \G({m^2-u\over 2}+1)
\G({m^2-s\over 2}+1)/[\Gamma({t\over 2} + 2) \Gamma({u - m^2 \over 2})
\Gamma({s - m^2 \over 2})]
\nonumber
\eea
and
\be
\ba
K_R^1 = \eta_R.\eta'_R \zeta_R.\zeta'_R, \quad K_R^2 =
\zeta_R.\zeta'_R \eta_R.Q_R \eta'_R.Q_R\\
K_R^3 = \zeta_R.\eta_R \zeta'_R.\eta'_R, \quad K_R^4 =\zeta_R.\eta'_R
\zeta'_R .\eta_R \\
\ea
\label{2.6}
\ee
\be
\ba
K_L^1 = \eta_L.\eta'_L \zeta_L.\zeta'_L, \quad K_L^2 =
\zeta_L.\zeta'_L \eta_L.Q_L \eta'_L.Q_L\\
K_L^3 = \zeta_L.\eta_L \zeta'_L.\eta'_L,
\quad K_L^4 =\zeta_L.\eta'_L \zeta'_L .\eta_L \\
\ea
\label{2.6a}
\ee
In the above $\V_B^{(-)}$ indicates that the corresponding vertex
operator carries charge $-Q_R, -Q_L$, ensuring charge
conservation. Since our probe here is neutral, the left and right
moving charges of the EBPS state cannot change in the scattering
process.

In order to start comparing with the classical scattering, let us work
in the rest frame of the initial EBPS state (particle 1). We use the
following notation:
\be
k_1 =(m, \vec 0), \, k_2= (w, \vec k), \, k_3 = (E', \vec k - \vec
k'), \, k_4 = (w', \vec k'); \, \vec k. \vec k' \equiv w w' \cos
\theta
\label{restframe}
\ee
With this, $t= - 4 w w' \sin^2 {\theta \over 2}$, $s = m^2 + 2 wm$ and
$ u = m^2 - 2 mw + 4 w w' \sin^2{\theta \over 2} $, where by momentum
conservation, $w' = w(1 + 2(w/m) \sin^2(\theta/2))^{-1}$. It follows
that for $w<<m_P$ (which implies $w << m$)
\be
A_1(s,t,u) = \pi \alpha' \frac{m^2}{4} {\rm cosec}^2
\frac{\theta}{2}[1 + o(w/m)]
\label{avalue}
\ee
and that the all terms in \eq{2.4} other than the one containing the
product $K_R^1 K_L^1$ are either down by $w/m_P$ or they are down by
$(m_P/m)^2$. Thus we get
\be
\M(1,2,3,4) = \pi \alpha' \frac{m^2}{4}
{\rm cosec}^2 \frac{\theta}{2}[K_1 +
o(w/m) + o(m/m_P)^2 ]
\label{mnlone}
\ee
with
\be
K_1 = K_R^1 K_L^1
\label{kone}
\ee
In equations \eq{avalue} and \eq{mnlone} we have reinstated
$\alpha'(=1/m_P^2)$.

\vspace{2 ex}
\underbar{Case $N_L > 1$}:
The holomorphic part of the vertex operator for the EBPS state,
$V_B(z)$, remains as above, while there are now many choices of the
antiholomorphic part, corresponding to the various ways a state with a
given $N_L>1$ can be constructed. Let us choose a basis $\bar
V_B^{(r)}(\z)$ for these vertex operators. In general these will
satisfy an OPE:
\be
\bar V_B^{(r)}(\z) \bar V_B^{(s)}(\w)  =
\frac{Z^{rs}}{(\z - \w)^{2 N_L}}
\label{2.7}
\ee
In the following (see the normalization convention in Appendix C) we
will use an orthonormal basis of vertex operators so as to make
$Z^{rs} = \delta_{rs}$. We will denote the full vertex for the
EBPS state as
\be
\V_B(\zeta_R; r; k; z, \z) =
V_B(\zeta, k,z) \bar V_B^{(r)}(\z) \exp[iQ_R.x_R + iQ_L.x_L + ik.x(z,\z)]
\label{2.8}
\ee
Examples are: $\bar V_B^{(r)}(\z) = \zeta_{L,i} \del^2_\z x^i$ or
$= \zeta_{L,ij} \del_\z x^i \del_\z x^j$ for $N_L =2$.

Let us denote the initial and final polarizations of the EBPS state as
$(\zeta_R, r)$ and  $(\zeta'_R, s)$. The four-point amplitude is now
given by (Appendix B)
\be
\ba
\M(1,2,3,4) \equiv \int  d^2 z_3 \, \langle c\bar c(z_1)
c\bar c (z_2) c\bar c(z_4) \\
\V_B(\zeta_R; r; k_1; z_1, )
\V_M(\eta_R;\eta_L; k_2; z_2, \z_2)
\V_B^{(-)}(\zeta'_R; s; k_3; z_3, \z_3)
\V_M(\eta'_R; \eta'_L; k_4; z_4, \z_4)
\rangle  \\
= A_1(s,t,u) [ K_1 + o(w/m_P) + o(m/m_P)^2 ]
\ea
\label{2.9}
\ee
$K_1$ is once again defined by \eq{kone} with $K_L^1$ this time
given by
\be
K_L^1 = \eta_L.\eta'_L Z^{rs}
\label{2.10}
\ee
The low energy (and $m>>m_P$) limit of the amplitude $\M(1,2,3,4)$ is
given once again by the equation \eq{mnlone}.

\vspace{2 ex}
\noindent{\bf No hair}:
Note that the internal polarization of the probe and the EBPS state
are decoupled in the factor $K_1$ for both $N_L = 1$ and $N_L >1$.  By
way of contrast, terms like $K_R^3$ (Eqn. \eq{2.6}) have inner
products between the internal polarizations of the probe and those of
the EBPS state.  Amplitudes involving the term $K_R^3$ can be used, by
sending in probes in suitable internal state of polarization, to
measure the state of internal polarization of the EBPS state, in
contradiction to the no-hair theorem. However, as we have seen above,
such terms disappear in the low energy limit.

\vspace{2 ex}
\noindent{\bf Inelastic threshold}: It is easy to deduce the
inelastic threshold, namely $w'_{cr}$, such that for $w> w'_{cr}$
the probe can excite the EBPS state to the next higher mass state
with the same charges $Q_R, Q_L$ (remember that the probe, being
neutral, cannot change these charges). The general mass formula
\be
m^2 = |Q_R|^2 + 2 N_R -1 = |Q_L|^2 + 2 N_L - 2
\label{2.12}
\ee
(BPS condition is $N_R = 1/2$) says that the next higher mass state
(mass $m'$) above the EBPS state is the one obtained by letting $N_R
\to N_R +1, N_L \to N_L + 1$, so that
\be
m^{\prime 2} = m^2 + 2
\label{2.13}
\ee
The state obtained by  $N_R \to N_R + n, N_L \to N_L + n$ has a mass
$m^{(n)}$ where
\be
[m^{(n)}]^2  = m^2 + 2n
\label{2.14}
\ee
It is a simple exercise in relativistic kinematics to show that the
state $m'$ cannot be excited unless the probe has a frequency $w \ge
w'_{cr}$\footnote{We have explicitly checked, for $w \ge w'_{cr}$,
that the tree level amplitude for the process EBPS $+$ probe $\to$
non-BPS (mass $m'$) $+$ probe is non-zero.}  (in the
rest frame of the original EBPS state) where
\be
w'_{cr} = 2/(m \alpha')
\label{2.15}
\ee
where we have reinstated $\alpha'$.

\section{Comparison between String Scattering and Classical Scattering}

\noindent{\bf Scattering amplitude}:
The string amplitude $\M(1,2,3,4)$ leads to the following scattering
cross-section (Appendix C) in the rest frame of `1':
\be
d\sigma/d\Omega = \frac{E_4^2}{E_2^2 E_1 E_3}
|\M(1,2,3,4)|^2 (2 \pi)^{-2}
\label{3.1}
\ee
and $\sigma \equiv \int \sin\theta\, d\theta\, d\phi \,
d\sigma/d\Omega$, where $\theta, \phi$ are the relative angles between
the vectors $\vec k_4$ and $\vec k_2$. On the other hand, the
cross-section in the classical scattering is given by
\be
d\sigma/d\Omega = | f(\theta) |^2
\label{3.2}
\ee
If the geometry around the EBPS state is indeed correctly reproduced
by the classical black hole solution, then the two scattering
cross-sections should agree. Therefore we must have
\be
f(\theta) = \frac{E_4}{2 \pi E_2 \sqrt{E_1E_3}} \M(1,2,3,4)
\label{3.3}
\ee
upto a constant phase factor.

Now the low energy $w/m_P \to 0$ limit of \eq{1.11}\ and \eq{mnlone}
are, respectively,
\be
f(\theta) = \alpha'{ m \over 8} {\rm cosec}^2 {\theta \over 2}
\label{flow}
\ee
and
\be
\M(1,2,3,4) = \alpha' {\pi m^2 \over 4} {\rm cosec}^2 {\theta \over 2}
[1 + o(m/m_P)^2]
\label{mlow}
\ee
In the last equation we have used the fact that $K_1$ gives rise to 1
when we sum and average $|K_1|^2$ in \eq{3.1} over final and initial
polarization states respectively. Noting that to leading order in
$w/m_P$, $E_1 =E_3 = m, \quad E_2 = E_4 = w$, we find that the right
hand side of
\eq{3.3} is
\be
\alpha' {m \over 8} {\rm cosec}^2 {\theta \over 2} [ 1 + o(m/m_P)^2]
\ee
which {\em precisely} agrees with  \eq{flow} in the limit $m_P/m
\to 0$.

We should note that in \eq{1.11} there are correction terms of the
form $w m_0 = 4(w/m_P)(m/m_P)$ which implies that we must {\em first}
expand to leading order in $w/m_P$ before taking the large $m/m_P$
limit; alternatively we should define $w<< m_P (m_P/m)$ as the
appropriate low-energy limit.

\vspace{2 ex}
\noindent{\bf Disappearance of hair}: As we remarked after \eq{2.10},
terms in the scattering amplitude which potentially constitute a
measurement of the internal polarizations (`hair') of the EBPS state
disappear in the same limit as discussed above.

\vspace{2 ex}
\noindent{\bf Thresholds}:
It is trivial to see that the inelastic thrsehold is given by
\be
w'_{cr} = 2/{\alpha' m} = \ch/(2 m_0)
\ee
which agrees with the absorption threshold \eq{1.12} in the limit $\ch
\to 1$, or equivalently $N_L \to N_{\rm max} \equiv 1 + \alpha'
m^2/4$ (note that by \eq{2.2} or \eq{2.12} this is the maximum
possible value of $N_L$ for a given mass).  Since the degeneracy of
the BPS state essentially comes from the number of left-moving
oscillators (in the right sector the BPS condition forces $N_R =1/2$),
therefore $N_L = N_{\rm max}$ is also the limit when the degeneracy of
the EBPS state, for a given mass, is maximized.

\section{Concluding Remarks}

We have seen that the geometry around the large mass EBPS state is the
same as that of the classical extremal black hole solution as seen by
a masslass scalar probe. The agreement has been verified to the
leading order in the low energy expansion. It is interesting to note
that a curved space sigma model description of string theory which is
normally supposed to describe a `string condensate' of massless modes
represents a very high mass elementary string state in our scattering
experiment.

It is also interesting to observe that the low energy limit ($w <<
m_P$) restores the no-hair theorem, since the terms capable of
measuring the internal polarization state of the EBPS state (which is
responsible for its degeneracy, given the mass and the charges) drop
out in this limit.  The interpretation of such terms in the case $w
\approx m_P$ or in the case where the leading term in \eq{mnlone} is
made to disappear by choosing the initial and final polarization of
the probe to be orthogonal remains an interesting open problem.

Another intriguing result that we found is the emergence of the $N_L
\to N_{\rm max}$ limit where the classical absorption threshold and the
inelastic threshold of the string scattering agree. In a sense the
test provided by this agreement is more stringent than the agreement
of the scattering amplitude, since absorption is a rather essential
feature of black hole geometry as against scattering in conventional
central fields of force. Note that we need $N_L$ to be large also for
the agreement \cite{Sen} between the degeneracy of the EBPS state and
the Beckenstein-Hawking entropy of the classical black hole (the
degeneracy formulae do not hold for $N_L$ of order 1).  We have
already remarked that $N_L = N_{\rm max}$ corresponds to the maximum
degeneracy of the EBPS state for a given mass. The existence of a
limit where the two thresholds agree gives a rather simple picture in
that limit of what happens when matter falls into the black hole. The
black hole absorbs the energy and gets excited to a higher vibrational
state of the string. This higher mass state is a non-BPS state and
hence it decays according to standard string theory back to the BPS
state. It would be extremely interesting to see under what circumstances
such a decay might possibly correspond to Hawking radiation. Work
in this direction is in progress.

\vspace{3 ex}
Acknowledgement: We would like to thank A. Dhar and A. Sen for many
useful discussions. G.M. would like to thank R.S. Bhalerao, S.S. Jha,
S.M. Roy, K.V.L. Sarma and N. Ullah for discussions regarding
scattering theory for the modified Coulomb problem.

\vspace{3 ex} Note added: After this paper was completed,
the paper \cite{Callan} was pointed out to us by A. Sen which has some
overlap with this work.

\appendix

\section{Derivation of the scattering amplitude $f(\theta)$}

The following sum over partial waves is well-known from the theory of
scattering in a Coulomb potential \cite{Schiff}:
\be
\ba
\sum_{l=0}^\infty  (2 l + 1) i^l \exp[i\eta_l] P_l (\cos \theta)
\sin[ w  \tilde \rho  + \eta_l - l \pi/2]/(w \rho) \\
\,\sim \exp[ i w (z - m_0 \, \ch \, \ln w(\rho - z))] + \rho^{-1}
f_c(\theta) \exp[ i w \tilde \rho]
\ea
\label{a.1}
\ee
where
\bea
\eta_l = {\rm arg} \Gamma( l +1 - ia'), \; l=0,1, \ldots, \infty,
\nonumber
\eea
\be
\ba
f_c(\theta) = (1/2) m_0 \ch \,{\rm cosec}^2 {\theta \over 2}
\exp i \varphi,\\
\varphi = m_0 w \, \ch \ln \sin^2 (\theta/2) + 2 \eta_0
\ea
\label{a.2}
\ee
and $a'$ and $\tilde \rho$ are as defined in \eq{1.7} and in \eq{1.9}.
The logarithmic corrections to the phases of the incident and the
outgoing waves are characteristic of any potential with a $1/\rho$
fall-off at infinity. The above equation is meant to be true at
asymptotically large distances $\rho$.

Since we have a modified Coulomb potential \eq{1.5}, our phase shifts
$\delta_l$, determined in \eq{1.9}, differ from $\eta_l$. However,
for small $m_0 w$, it is easy to see that
\be
l' =  l - (m_0 w)^2 / (2 l + 1) + o(m_0 w)^4
\nonumber
\ee
so that
\be
\delta_l = \eta_l + \Delta_l, \Delta_l = {1 \over 2 l +1}[ o(m_0 w)^2]
\label{a.3}
\ee
In order to arrive at a sum over partial waves analogous to \eq{a.1}\
for our present problem we do the following manipulations \cite{Schiff}:
\be
\ba
(w \rho)^{-1}
\sum_{l=0}^\infty  (2 l +1) i^l \exp[i\delta_l] P_l (\cos \theta)
\sin[ w  \tilde \rho  + \delta_l - l \pi/2] \\
=  (2 i w \rho)^{-1} \sum_l (2 l + 1) i^l \exp[i \eta_l] \bigg(
[\exp(2 i \Delta_l) - 1] \exp[iw\tilde \rho] \exp i
[\eta_l - l \pi/2] + \\
\,~~~~~~~~~~~~~~~~~~~~~~~~
\sin[ w\tilde \rho + \eta_l - l \pi/2] \bigg) P_l
(\cos \theta) \\
\sim \exp[ i w (z - m_0 \, \ch \, \ln w(\rho - z))] + \rho^{-1}
[f_c(\theta) + f_m(\theta)] \exp[ i w \tilde \rho],
\ea
\label{a.4}
\ee
\be
f_m(\theta) = {1 \over w} \sum_l (2 l + 1) \exp[2 i \eta_l]{
\exp[2 i \Delta_l] - 1 \over 2 i} P_l(\cos \theta)
\label{a.5}
\ee
Using \eq{a.3} it is easy to see that the sum in \eq{a.5} is
convergent and $f_m(\theta) = m_0 \times [o(m_0 w)]$. Noting further
that $\varphi = o(m_0 w)$ in \eq{a.2}, we find that \eq{a.4} reduces
to the right hand side of \eq{1.10} with $f(\theta)$ given in
\eq{1.11}. In order that $M_w$, given by \eq{1.4} and \eq{1.9},
reduces to the left hand side of \eq{a.4} we need to choose
\bea
C_{wl} = \sqrt{4\pi(2 l + 1)}i^l \exp[2 i \delta_l]/ w
\nonumber
\eea
This concludes the derivation of \eq{1.11}. The additional $o(w/m_P)$
correction has been added in \eq{1.11} because the metric itself
can get corrected to the next order in $\alpha'$, leading to such
corrections.

\section{Details of String Scattering}

We reproduce only the essential steps here. The basic method is the
same as in \cite{FMS}.

\vspace{2 ex}
\underbar{Case $N_L=1$}

The integrand in \eq{2.4} reduces to a product $R\, L\, K$ where
\bea
R = \langle cV_B(z_1) \exp[iQ_R.x_R(z_1)] cV_M(z_2)
V_B(z_3) \exp[- i Q_R.x_R(z_3)] cV_M(z_4) \rangle
\nonumber
\eea
\bea
L = \langle  \bar c \bar V_B(\z_1) \exp[iQ_L.x_L(\z_1)]
\bar c \bar V_M(\z_2)
\bar V_B(\z_3) \exp[- i Q_L.x_L(\z_3)]
\bar c \bar V_M(\z_4) \rangle
\nonumber
\eea
\bea
K = \langle \prod_i \exp[i k_i.x(z_i, \z_i)] \rangle
\nonumber
\eea
Before writing down the expressions for $R$ and $L$ let us, as in
\cite{FMS}, use the $SL_2(C)$ invariance of the four-point amplitude
to choose $z_1=0, z_2 =1$ and $z_4 \to \infty$. We will denote $z_3$
by $z$. The correlation functions in this notation evaluate to
\bea
R &=& (-z)^{- m^2 -2} \bigg[\zeta_R.\zeta'_R \eta_R.\eta'_R (1 -
k_2.k_4) + z \, k_2.k_4 \zeta_R.\eta_R \zeta'_R. \eta'_R
\nonumber\\
\,&-& {z\over 1-z} k_2.k_4 \zeta_R.\eta'_R \zeta'_R.\eta_R
+ {z^2\over 1-z}
\zeta_R.\zeta'_R \eta_R.Q_R \eta'_R.Q_R \bigg] \nonumber\\
\label{rvalue}
\eea
\bea
L = (-\z)^{-m^2 -2}\bigg[ \zeta_L.\zeta'_L \eta_L. \eta'_L &+& \z^2
\zeta_L.\eta_L \zeta'_L.\eta'_L + ({\z\over 1 - \z})^2
\zeta_L. \eta'_L \zeta'_L .\eta_L \nonumber\\
\, &-& {\z^2 \over 1 - \z} \zeta_L.\zeta'_L \eta_L.Q_L \eta'_L.Q_L
\bigg]\nonumber\\
\label{lvalue}
\eea
\bea
K = |z|^{2 k_1.k_3} |1 - z|^{2 k_2.k_3}
\nonumber
\eea
We now integrate $R \, L \, K$ using the formula \cite{GSW},
\be
\ba
\int d^2 z z^{a + n_1} (1-z)^{b + n_2} \z^{a+n_3} (1 - \z)^{b + n_4}
 = - \sin \, \pi b \,\times \\
\,~~~~~~~~~~~~~~~~~~~~~ B(a+n_1 + 1, b + n_2 + 1) \, B(- (a+b+ n_3 +
n_4 +1), b + n_4 + 1)
\ea
\label{integral}
\ee
We put $a = k_1.k_3 - m^2 - 2 = -t/2 - 2$ and $b = k_2.k_3 = (m^2 -
u)/2$. By using $\Gamma(z+1) = z \Gamma(z) $ repeatedly we get
\eq{2.4}.

\underbar{Case $N_L>1$}

The factors $R$ and $K$ remain the same. Our strategy for $L$ is as
follows.  It is not difficult to see that the $n_3 = n_4 =0 $ term
always gives rise to the same kinetic factor ($A_1(s,t,u)$) as in the
case $N_L =1$. Does the pattern that other values of $n_3, n_4$ give
rise to integrals which are down either by $w/m_P$ or by $(m_P/m)^2$
persist? Indeed it does. Proof: the entire set of values of $n_3, n_4$
($n_3\ge 0, n_4 \le 0$) can be classified in terms of the ratio
\bea
f(n_3, n_4) = {B(-(1 + a + b + n_3 + n_3), b + n_4 + 1)
\over B(-(1+ a +b), b+1)}
\nonumber
\eea
as follows:
\bea
(n_3, n_4) = (0,0) : \quad &f& = 1 \nonumber\\
n_3 \in \{0,1\}, \, n_4 \le -1 : \quad &f& = o(1/mw) \nonumber\\
(n_3, n_4) = (1,0) : \quad &f& = o(1/mw) \nonumber\\
n_3 \ge 2, n_4 =0 : \quad &f& = o(w/m) \nonumber\\
n_3 + n_4 \ge 1, n_4 \le -1 : \quad &f& = o(1/m^2) \nonumber\\
n_3 + n_4 \le 0, n_3 \ge 2 : \quad &f& = o(w/m) \nonumber
\label{listf}
\eea
It is not too difficult to see that contractions corresponding to the
second and third lines do not appear in $L$ at all. The crucial thing
to note is that there is no other term than the top line which
survives the $w/m_P \to 0, m_P/m \to 0$ limit. This proves \eq{2.13}.

\section{Normalization Convention for S-matrix}

We discuss here the relative coefficient between $S$-matrix elements
and string scattering amplitudes.  Suppose
\be
\langle 3,4 | -i S | 1,2 \rangle = C_0 \M(1,2,3,4)
[\prod_i 2 w_i \Omega]^{-1/2} (2 \pi)^4 \delta^{(4)} (\sum_i k_i)
\ee
\be
\langle 3 | -i S | 1,2 \rangle = C_0 \M(1,2,3)
[\prod_i 2 w_i \Omega]^{-1/2} (2 \pi)^4 \delta^{4} (\sum_i k_i)
\ee
where $\M(1,2,3,4)$ is as defined in \eq{2.4} and $\M(1,2,3) = \langle
c\bar c V_1 c\bar c V_2 c\bar c V_3 \rangle$. The vertex operators are
normalized to satisfy an OPE: $V(z) V(w) = |z - w|^{-4}$.  In the
above equations we follow the convention and notation of
\cite{Lee}, where $\Omega$ is the volume of space. The $(2 \pi)^4
\delta^{(4)}(\sum_i k_i)$ stands for the integral $\int d^4 x\, \exp[i
k.x]$ over a space box of volume $\Omega$ and a time interval $T$. The
constant $C_0$ can be fixed by demanding tree-level unitarity in some
simple example. In these conventions $C_0$ turns out to be $4$.

With the above convention, the relation between the $S$-matrix element
and the scattering cross-section (in the rest frame of particle `1') is
\cite{Lee}
\be
d\sigma = \sum_{3,4} | < 3,4 | S | 1, 2> |^2 \times \Omega/(v_2 T)
\ee
where $v_2$ is the velocity of particle $2$ in the rest frame of $1$.
Using the standard manipulations of squaring a delta-function etc., it
is straightforward to arrive at \eq{3.1}.


\end{document}